\documentclass[12pt]{article}
\usepackage{amssymb}
\usepackage{amsfonts}
\usepackage{amsmath}
\usepackage[OT4]{fontenc}

\newtheorem{thm}{Theorem}[section]
\newtheorem{fact}{Fact}[section]
\newtheorem{lem}{Lemma}[section]

\newtheorem{col}{Corollary}[section]

\newenvironment{pf}{\noindent \textit{Proof.}}

\newenvironment{pf*}[1]{\noindent\textit{#1} }{}

\long\def\symbolfootnote[#1]#2{\begingroup%
\def\thefootnote{\fnsymbol{footnote}}\footnote[#1]{#2}\endgroup}

\newcommand{\qed}{$\square$}
\newcommand{\ints}{\int\limits}
\newcommand{\sgn}{\textrm{sgn}}

\def\comp{{\rm comp}}
\def\quant{{\rm quant}}
\def\rand{{\rm rand}}

\def\dc{{\rm d}}

\begin{document}
\begin{center}
{\Large \textbf {Randomized and Quantum Solution\\ of Initial-Value
Problems
\\\vspace{0.4cm}for Ordinary Differential Equations of Order $k$}\symbolfootnote[1]{$\;$This research was partly supported by AGH grant No. 10.420.03}\vspace{1.5cm}}

Marek Szcz\c{e}sny \vspace{.5cm}

\textit{Faculty of Applied Mathematics, AGH University of Science
and Technology,\\ Al. Mickiewicza 30, 30-059 Cracow, Poland,\\
E-mail:} \texttt{szczesny@uci.agh.edu.pl}
\end{center}
\vspace{1.0cm}
\begin{abstract}
We study possible advantages of randomized and quantum computing over
deterministic computing for scalar initial-value problems for ordinary
differential equations of order $k$. For systems of equations of the
first order  this question has been settled modulo some details in
\cite{Kacewicz05}. A speed-up over deterministic computing shown
in~\cite{Kacewicz05} is related to the increased regularity of the
solution with respect to that of the right-hand side function. For a
scalar equation of order $k$ (which can be transformed into  a special
system of the first order), the regularity of the solution is
increased by $k$ orders of magnitude. This leads to improved
complexity bounds depending on $k$ for linear information in the
deterministic setting, see~\cite{Szczesny05}. This may  suggest that
in the randomized and quantum settings a speed-up  can also be
achieved depending on $k$.

We show in this paper that a speed-up dependent on $k$ is not possible
in the randomized and quantum settings. We establish lower complexity
bounds, showing that the randomized and quantum complexities remain at
the some level as for systems of the first order, no matter how large
$k$ is. Thus, the algorithms from~\cite{Kacewicz05} remain (almost)
optimal, even if we restrict ourselves to a subclass of systems
arising from scalar equations  of order $k$.
\end{abstract}

\section{Introduction}

In the previous paper~\cite{Szczesny05} we established complexity
bounds for scalar initial-value problems for ordinary differential
equations of order $k$ in the deterministic worst-case setting. We
showed for instance that if the right-hand side function $g$ depends
only on the solution function (not on the derivatives of the
solution), then the $\varepsilon$-complexity for linear information is
equal to $\Theta\left(\left(1/\varepsilon\right)^{1/(r+k)}\right)$,
where $r$ denotes the regularity of $g$. Hence, the order $k$ of the
equation contributes significantly to the $\varepsilon$-complexity. In
this paper, we study the dependence of the $\varepsilon$-complexity on
$k$ in the randomized and quantum settings. So far, the complexity in
the randomized and quantum settings was considered for systems of
equations of the first order, see~\cite{Kacewicz05}.  For right-hand
side functions with $r$ continuous derivatives, the
$\varepsilon$-complexity is essentially of order
$\left(1/\varepsilon\right)^{1/(r+\varphi)}$, where $\varphi=0$ in the
deterministic setting, $\varphi=1/2$ in the randomized setting, and
$\varphi=1$ in the quantum setting (for details see Section~3). A
speed-up for systems of equations over deterministic computing is thus
by~$1/2$ in the exponent in the randomized case, and by~$1$ in the
quantum case. Since a scalar equation of order~$k$ can be written as a
special system of $k+1$ equations of the first order, the upper bounds
from~\cite{Kacewicz05}~remain valid in this case.

Intuitively, for scalar equations of order $k$, due to the increased
regularity of the solution, one might expect a better speed-up over
deterministic computing related to $k$. As we mentioned above, an
improvement dependent on $k$ is achieved  in the deterministic
worst-case setting by passing from the standard to linear (integral)
information. One may hope that proper randomized or quantum
approximation of the integrals involved in the computations will lead
to algorithms with improved error bounds dependent on~$k$.

We show in this paper that such an improvement is not possible, and
a speed-up in the randomized and quantum settings is independent of
$k$. We establish lower complexity bounds, showing that the
complexity is
$\Omega\left(\left(1/\varepsilon\right)^{1/(r+1/2)}\right)$ in the
randomized setting, and
$\Omega\left(\left(1/\varepsilon\right)^{1/(r+1)}\right)$ in the
quantum setting, no matter how large $k$ is. Hence, the algorithms
defined in~\cite{Kacewicz05} remain almost optimal even if we
restrict ourselves to special systems arising from scalar problems
of order~$k$.

The paper is organized as follows. We first  introduce necessary
definitions in the three settings. Then we recall  for further
comparison known complexity bounds  for  initial-value problems.
Main results, lower bounds on the $\varepsilon$-complexity in the
randomized and quantum settings, are shown in Section~\ref{section
Lower bound}.

\section{Problem definition}

We  consider the complexity of a problem in the following form
\begin{equation}
\left\{\begin{array}{l}u^{(k)}(x)=g\left( x,u(x),u^{\prime
}(x),\ldots ,u^{(q)}(x)\right) ,\qquad
x\in \lbrack a,b],\\
u^{(j)}(a)=u_{a}^{j},\quad j=0,1,\ldots ,k-1,%
\end{array}
\right. \label{general problem}
\end{equation}
where $0\leq q<k$, $g:[a,b]\times \mathbb{R}^{q+1}\rightarrow \mathbb{R}$, $%
u:[a,b]\rightarrow \mathbb{R}$ ($a<b$).

\noindent  For $r \geqslant 1$ and given positive numbers
$D_0,\ldots, D_r$, we consider the class of right-hand side
functions $g$ defined by
\begin{eqnarray}
&\mathcal{G}^r=\left\{g\;|\;g\in C^{(r)}([a,b]\times
\mathbb{R}^{q+1}),\quad |\partial^jg(x,y)|\leqslant D_{j},\quad
\textrm{for}
\right.&\nonumber\\
& x\in [a,b], \quad  y\in \mathbb{R}^{q+1},\quad j=0,1,\ldots r
\big\},& \label{class G}
\end{eqnarray}
where $\partial^jg$ represents all partial derivatives of order $j$
of $g$.

 Instead of~(\ref{general problem}) we can write an equivalent system
of differential equations of order~1 of the form:
 \begin{equation}
\mathbf{u}'(x)=\left[\begin{array}{c} u'_0(x)\\
u'_1(x)\\
\vdots \\u'_{k-1}(x)\\u'_{k}(x)
\end{array}\right]=\left[\begin{array}{c} 1\\
u_2(x)\\
 \vdots \\
 u_{k}(x)\\
 g(u_0(x),u_1(x),\ldots,u_{q+1}(x))
\end{array}\right]=\mathbf{g}(\mathbf{u}(x)),\;x\in[a,b],
\label{system of ODEs}
\end{equation}
with a initial conditions
\begin{equation}
\mathbf{u}(a)=\left[ a, u^0_a, \ldots ,u^{k-1}_a \right]^T.
\label{initial conditions for system of ODEs}
\end{equation}

\noindent Then the solution $u(x)$ of (\ref{general problem})
corresponds to the function $u_1(x)$.

 Before we start analyzing the complexity of~(\ref{general
problem}), we remind  basic notions connected with deterministic,
randomized and quantum computations. We are interested in finding a
bounded function $l=l(x)$ that approximates the solution
of~(\ref{general problem}). The construction of $l$ is based on a
certain information about the right-hand side function $g$. In the
deterministic setting, we usually consider standard information, in
which we  compute values of $g$ or its partial derivatives at some
points, or  linear information in which we know values of linear
functionals of $g$.

In the randomized  setting  the values of $g$ or its partial
derivatives can be computed at randomly chosen points. In the
quantum setting, information about $g$ is gathered by applications
of a quantum query for $g$. The reader is referred
to~\cite{Heinrich} for a detailed explanation what a quantum query
is.

 To get an approximate solution $l(x)$, we use an algorithm $A$,
 which is a mapping from the information space into the space of
 bounded functions.

\noindent In the deterministic setting, the {worst-case error} of an
algorithm $A$  in class $\mathcal{G}^r$ is defined by
\begin{equation}
e^{\textrm{worst}}(A ,\mathcal{G}^r)=\sup\limits_{g\in
\mathcal{G}^r}\sup\limits_{x\in [a,b]}|u(x)-l(x)|.
\end{equation}
In the randomized and quantum settings  an approximation obtained is
random. Letting $\left(\Omega,\Sigma,\textbf{P}\right)$ be a
probability space, an algorithm $A$ provides us with  an approximate
solution $l^\omega$, where $\omega \in \Omega$.

\noindent The error of the algorithm $A$ at $g$ is defined by
\begin{equation}
e^\omega(A,g)=\sup_{x\in[a,b]}|u(x) - l^{\omega}(x)|
\end{equation}
(we assume that $e^\omega(A,g)$ is a random variable for each $g \in
\mathcal{G}^r$). The {error of} $A$ {in the class} $\mathcal{G}^r$
in the randomized setting is defined by
\begin{equation}
e^{\textrm{rand}}(A,\mathcal{G}^r)=\sup_{g\in
\mathcal{G}^r}\left(\textbf{E}(e^\omega(A,g))^2\right)^{1/2},
\end{equation}
and in the quantum setting by
\begin{equation}
e^{\textrm{quant}}(A,\mathcal{G}^r)=e^{\textrm{quant}}(A,\mathcal{G}^r,\delta)=\sup_{g\in\mathcal{G}^r}\inf\{\alpha
\;|\; \textbf{P}(e^\omega(A,g)>\alpha) \leq \delta \}.
\end{equation}
 The number $\delta\in (0,1/2)$ denotes here
the failure probability. It is often assumed to be $1/4$. The
 success probability can then be increased  by taking a
median of a number of  repetitions of  an algorithm $A$
(see~\cite{Heinrich}).

By the  cost in the deterministic, randomized and quantum settings,
we mean a number of subroutine calls for $g$, which is used by an
algorithm $A$. Thus, in the deterministic and randomized settings,
the cost is equal to number of evaluations of $g$ or its partial
derivatives, while in the quantum setting it is a number of quantum
query calls. We will denote the cost in the respective setting by
$\mathrm{cost}^{\textrm{worst}}(A)$,
$\mathrm{cost}^{\textrm{rand}}(A)$ or
$\mathrm{cost}^{\textrm{quant}}(A)$.

 For
any $\varepsilon>0$, by the { $\varepsilon$-complexity} of the
problem we mean the minimal cost sufficient  to solve the problem
with error no larger  than $\varepsilon$, where the minimum is taken
over all algorithms solving the problem
\begin{eqnarray}
\comp(\mathcal{G}^r,\varepsilon ) &=&\min_A \left\{
\mathrm{cost}(A)\;|\; e(A,\mathcal{G}^r)\leqslant \varepsilon
\right\}.
\end{eqnarray}
To denote the complexity in the deterministic, randomized or quantum
settings, we will use a suitable superscript: "$\textrm{worst}$",
"$\textrm{rand}$" or "$\textrm{quant}$". Additionally to denote
different types of information used  in the deterministic setting,
we will use  indices: "$\textrm{worst-st}$" and
"$\textrm{worst-lin}$" for standard and linear information,
respectively.

\section{Upper complexity bounds}
In this section we briefly recall known complexity bounds for scalar
equations of order~$k$, as well as those for systems of the first
order.

In~\cite{Kacewicz05}, Kacewicz dealt with systems of equations of
the first order of the form
\begin{equation}
z'(t)=f(z(t)),\quad t\in [a,b],\quad z(a)=\eta, \label{autonomous
equation kacewicz}
\end{equation}
where $f:\mathbb{R}^d\rightarrow \mathbb{R}^d$ and $\eta \in
\mathbb{R}^d$. He  considered the H\"older class of functions
\begin{eqnarray}
&\mathcal{F}^{r,\rho}=\big\{f:\mathbb{R}^d \rightarrow
\mathbb{R}^d\;|\: f\in C^{(r)}(\mathbb{R}^d),\quad |\partial^i f^j(y)|\leq D_i,\;i=0,1,\ldots,r,&\nonumber\\
& |\partial ^rf^j(y)-\partial^rf^j(z) |\leq H \|y-z\|^\rho,\;y,z\in
\mathbb{R}^d,\quad j=1,2,\ldots, d\big\},&
\end{eqnarray}
where $\rho\in(0,1]$.

\noindent It was shown in \cite{Kacewicz05} that the
$\varepsilon$-complexity is
\begin{equation}
\comp^{\textrm{rand}}(\mathcal{F}^{r,\rho},\varepsilon)=O
\left(\left(\frac{1}{\varepsilon}\right)^{1/(r+\rho+1/2-\gamma)}\right)
\label{Kacwicz rand holder}
\end{equation}
and
\begin{equation}
\comp^{\textrm{quant}}(\mathcal{F}^{r,\rho},\varepsilon)=O\left(
\left(\frac{1}{\varepsilon}\right)^{1/(r+\rho+1-\gamma)}\right)
\label{Kacewicz quant holder}
\end{equation}
with  an arbitrarily small positive parameter $\gamma$. (The constants
in the big-O notation  depend on $\gamma$.) These bounds are almost
optimal, i.e. they essentially match lower bounds on the complexity.

It is easy to see that the bounds above with $\rho=0$ hold for the
class $\mathcal{F}^r$, where
\begin{eqnarray}
&\mathcal{F}^{r}=\big\{f:\mathbb{R}^d \rightarrow
\mathbb{R}^d\;|\: f\in C^{(r)}(\mathbb{R}^d),\quad |\partial^i f^j(y)|\leq D_i,\;i=0,1,\ldots,r,&\nonumber\\
& \qquad y\in \mathbb{R}^d,\quad j=1,2,\ldots, d\big\}.& \label{class
Holder}
\end{eqnarray}
For systems~(\ref{autonomous equation kacewicz}) the
$\varepsilon$-complexity in the class $\mathcal{F}^r$ is thus equal to
\begin{equation}
\comp^{\textrm{rand}}(\mathcal{F}^{r},\varepsilon)=O
\left(\left(\frac{1}{\varepsilon}\right)^{1/(r+1/2-\gamma)}\right)
\label{Kacwicz rand}
\end{equation}
and
\begin{equation}
\comp^{\textrm{quant}}(\mathcal{F}^{r},\varepsilon)=O\left(
\left(\frac{1}{\varepsilon}\right)^{1/(r+1-\gamma)}\right).
\label{Kacewicz quant}
\end{equation}

Since equation (\ref{general problem}) can be  transformed into a
system of the first order (\ref{system of ODEs}),  upper complexity
bounds (\ref{Kacwicz rand}) and (\ref{Kacewicz quant})  are still
valid for scalar equations of order $k$ with $g\in\mathcal{G}^r$.
(Although function $\mathbf{g}$ has unbounded components, we can
consider an equivalent problem with a bounded right-hand side function
having bounded partial derivatives up to the $r$th order.)

The question that we deal with in this paper is whether these bounds
can be improved. We ask if a speed-up can be achieved dependent on $k$
due to the increased regularity of the solution. In some cases in the
deterministic worst-case setting such a speed-up dependent on $k$ can
indeed be shown. We have shown in~\cite{Szczesny05}  for standard
information that
\begin{equation}
\comp^{\textrm{worst-st}}(\mathcal{G}^r,\varepsilon )=\Theta
\left(\left( \frac{1}{\varepsilon }\right) ^{1/r}\right),
\end{equation}%
so that there is no dependence on $k$ in this case. However, if we
allow  linear information about right-hand side function,
 we can achieve a better result.  The use of integral
 information leads (for $q=0$) to the upper bound
\begin{equation}
\comp^{\textrm{worst-lin}}(\mathcal{G}^r,\varepsilon )=O
\left(\left( \frac{1}{\varepsilon }\right) ^{1/(r+k)}\right).
\end{equation}%
The complexity in this case significantly depends on $k$.

 Intuitively, one may expect  a speed-up
dependent on $k$ by replacing integrals in deterministic algorithms by
effective randomized or quantum approximations. In the next section we
show lower bounds on the complexity in both settings, which indicate
that this intuition turns out not to be correct.

\section{Lower complexity bounds in the randomized and quantum settings}
\label{section Lower bound} We show in this section  the main
results of this paper. We  prove the following lower bounds on
randomized and quantum complexity of equations of order $k$ with the
right-hand side function belonging to class $\mathcal{G}^r$.
Together with (\ref{Kacwicz rand}) and (\ref{Kacewicz quant}) they
show that the complexity of scalar equations~(\ref{general problem})
is independent of $k$.

\begin{thm} Let
$r\geqslant 1$. For an arbitrary $k$
\begin{equation}
{\rm comp}^{\rm rand}(\mathcal{G}^r,\varepsilon)=\Omega
\left(\frac{1}{\varepsilon}\right)^{1/(r+1/2)},
\end{equation}
and
\begin{equation}
{\rm comp}^{\rm quant}(\mathcal{G}^r,\varepsilon)=\Omega
\left(\frac{1}{\varepsilon}\right)^{1/(r+1)},%\log\frac{1}{\delta}
\end{equation}
where the constants in the ''$\Omega$'' notation depend only on the
class $\mathcal{G}^r$ and $k$.

 \label{theorem lower bounds}
\end{thm}

Before proving Theorem \ref{theorem lower bounds} we give some
auxiliary results. We  shall consider a subclass $\mathcal{G}^r_1$
of the class $\mathcal{G}^r$ given by
\begin{eqnarray}
&\mathcal{G}^r_1=\{ g:[a,b]\rightarrow \mathbb{R}\;|\: g\in
C^{(r)}\left(\left[a,b\right]\right)\; \sup\limits_{x\in
[a,b]}\left\vert g^{(j)}(x)\right\vert \leqslant D_j,\; j=0,\ldots
,r\}.&
  \label{klasa funkcji G_1}
\end{eqnarray}%
This class includes functions dependent on $x$ only, that is,
problem~(\ref{general problem}) for $g\in \mathcal{G}^r_1$ reduces
to  iterated integration. Since $\mathcal{G}^r_1\subset
\mathcal{G}^r$, we have that
\mbox{$\comp(\mathcal{G}^r_1,\varepsilon)\leqslant
\comp(\mathcal{G}^r,\varepsilon)$}.

We now show properties of auxiliary functions that will be used in
the proof of Theorem~\ref{theorem lower bounds}. Let
\begin{equation}
\psi_r(x):=\frac{1}{(r-1)!}\ints_{-1}^1(x-t)_+^{r-1}\sgn U_{r}(t)\dc
t\end{equation}
 be a perfect B-spline of degree $r$ with $r$ knots in
$(-1,1)$, see e.g.~\cite{Bojanov93}. Function $U_{r}$ is the
Tchebycheff polynomial of second kind, and $(x-t)_+^{r-1}$  a
truncated power function:
\begin{equation}
(x-t)_+^{r-1}:=\begin{cases}(x-t)^{r-1}&\textrm{for}\qquad
x\geqslant
t,\\
0 &\textrm{for}\qquad x<t.
\end{cases}
\end{equation}

\noindent For an interval $[c,d]$, we define by using  $\psi_r(x)$ a
function
\begin{equation}
\varphi_r([c,d],x):=\alpha
\left(\frac{d-c}{2}\right)^{r}\psi_{r+1}\left(\frac{2}{d-c}x-\frac{d+c}{d-c}\right),\quad
x\in[c,d],
\end{equation}
where $m_j:=\sup\limits_{x\in [-1,1]}|\psi^{(j)}_{r+1}(x)|$ and
$\alpha:=\min\limits_{j=0,1,\ldots,r}D_j/m_j$. From the properties of
perfect B-spline, the function $\varphi_r([c,d],\cdot)$ belongs to the
class $C^{(r)}([c,d])$, and
$\varphi^{(j)}_r([c,d],c)=\varphi^{(j)}_r([c,d],d)=0$ for
$j=0,\ldots,r$. In the sequel we will use the functions
$\varphi_r([c,d],\cdot)$ on intervals $[c,d]$ such that $d-c<2$ and
$[c,d]\subset[a,b]$. Note that $\varphi_r([c,d],\cdot)\in
\mathcal{G}^r_1$ for $d-c<2$ (if we define $\varphi_r([c,d],\cdot)=0$
beyond the interval $[c,d]$).

In the proof of Theorem \ref{theorem lower bounds} we will need the
value of the iterated integral of $\varphi_r([c,d],\cdot)$.
\begin{fact}
Let $m\in \mathbb{N}$, $m \geq 1$ and
\mbox{$\displaystyle\varphi_r^m:=\left(\frac{1}{2}\right)^{r+m}
\frac{\alpha}{(r+m)!}\ints_{-1}^1(1-t)^{r+m}\mbox{\rm sgn }
U_{r+1}(t)\dc t$}. Then
%\begin{enumerate}
%\item[a)]
\begin{equation}
\ints_{c}^{d}\ints_{c}^{y_{m-1}}\ldots\ints_{c}^{y_1}
\varphi_r([c,d],y_0)\dc y_0\,\dc y_1 \ldots \dc
y_{m-1}=(d-c)^{r+m}\varphi_r^m. \label{eq: B-spline 1}
\end{equation}
\label{fact B-spline 1}
\end{fact}
\begin{pf}
Changing variables $x_i=(2y_i-d-c)/(d-c)$, $i=0,1,\ldots,m-1$ in the
left-hand side of (\ref{eq: B-spline 1}), we get
\begin{eqnarray*}
L&=&\ints_{c}^d\ints_{c}^{y_{m-1}}\ldots\ints_{c}^{y_1}
\varphi_r([c,d],y_0)\dc y_0\,\dc y_1 \ldots \dc y_{m-1}\\
&=&\alpha\left(\frac{d-c}{2}\right)^{r+m}\ints_{-1}^1\ints_{-1}^{x_{m-1}}\ldots\ints_{-1}^{x_1}
\psi_{r+1}(x_0)\dc x_0\,\dc x_1 \ldots \dc x_{m-1}\\
&=&\alpha\left(\frac{d-c}{2}\right)^{r+m}\ints_{-1}^1\ints_{-1}^{x_{m-1}}\ldots\ints_{-1}^{x_1}
\frac{1}{r!}\ints_{-1}^1(x_0-t)_+^{r}\sgn U_{r+1}(t)\dc t\,\dc
x_0\,\dc x_1 \ldots \dc x_{m-1}.
\end{eqnarray*}
Changing the order of integration with respect to $t$ and $x_0$, and
 computing next the inner integral, we get
\begin{eqnarray*}
L&=&\alpha\left(\frac{d-c}{2}\right)^{r+m}\ints_{-1}^1\ints_{-1}^{x_{m-1}}\ldots\ints_{-1}^{x_2}
\frac{1}{(r+1)!}\ints_{-1}^1(x_1-t)_+^{r+1}\sgn U_{r+1}(t)\dc t\,
\dc x_1 \ldots \dc x_{m-1}.
\end{eqnarray*}
We proceed similarly for $x_i$, $i=1,\ldots,m-1$ to  get finally
\begin{eqnarray*}
L&=&\alpha\left(\frac{d-c}{2}\right)^{r+m}
\frac{1}{(r+m)!}\ints_{-1}^1(1-t)^{r+m}\sgn
U_{r+1}(t)\dc t \\
&=&(d-c)^{r+m}\varphi_r^m.
\end{eqnarray*}
\vspace*{-2cm}
\begin{flushright}
{\large{\qed}}%{\large $\blacksquare $ }
\end{flushright}
\end{pf}

For $n\in \mathbb{N}$, let $h:={(b-a)}/{n}$, and $a_i:=a+ih$ for
$i=0,1,\ldots,n$. We now consider  the following functions
\begin{equation} f_i(x):=\begin{cases}
\varphi_r([a_i,a_{i+1}],x) &\textrm{for}\quad
x\in[a_i,a_{i+1}]\\
0&\textrm{for}\quad x \notin[a_i,a_{i+1}]
\end{cases},\quad \textrm{for}\quad i=0,1,\ldots n-1.
\label{function f_i}
\end{equation}
Obviously, for $h<2$ functions $f_i$ belong to the class
$\mathcal{G}^r_1$ (see the  explanation after the definition of
$\varphi_r([c,d],\cdot)$).

The following lemma shows how to express the $k$-fold integrals of
the functions $f_i$.

\begin{lem}
Let functions $f_i$ be defined as above. Then for any $k \geq 1$ and
$n\geq 1$ we have
\begin{equation}
\ints_{a}^b\ints_{a}^{y_{k-1}}\ldots\ints_{a}^{y_1} f_i(y_0)\dc
y_0\,\dc y_1 \ldots \dc
y_{k-1}=h^{r+k}\sum_{m=1}^k\frac{1}{(k-m)!}\varphi_r^m (n-i-1)^{k-m}.
\label{ =C(i,n) varphi_r^m}
\end{equation}
%and
%\begin{equation}
%A\leqslant C(i,n)\leqslant B \label{A < C < B}
%\end{equation}
\label{lemat function f_i}
\end{lem}

\begin{pf}
To prove the lemma we shall show that

\begin{multline}
\ints_a^x\ints_a^{y_{k-1}}\ldots\ints_a^{y_1}f_i(y_0)\dc
y_0\ldots\dc
y_{k-2}\,\dc y_{k-1}=\\
=\begin{cases}
0 &\textrm{for}\quad x\in[a,a_i) \\
{\displaystyle \ints_{a_i}^x\ints_{a_i}^{y_{k-1}}\ldots\ints_{a_i}^{y_1} \varphi_r([a_i,a_{i+1}],y_0)\dc y_0\, \dc y_1\ldots \dc y_{k-1}}&\textrm{for}\quad x\in[a_i,a_{i+1}]\\
{\displaystyle \sum\limits_{m=1}^k \frac{h^{r+m}}{(k-m)!}\varphi_r^m
(x- a_{i+1})^{k-m}} &\textrm{for}\quad x \in (a_{i+1},b]
\end{cases}. \label{statement on integral of phi}
\end{multline}
We prove this by induction with respect to $k$. For $k=1$, we have
from Fact \ref{fact B-spline 1} with $m=1$ that
\begin{equation*}
\ints_a^xf_i(y) \dc y=\begin{cases}
0 &\textrm{for}\quad x\in[a,a_i) \\
{\displaystyle \ints_{a_i}^x \varphi_r([a_i,a_{i+1}],y)\,\dc y }&\textrm{for}\quad x\in[a_i,a_{i+1}]\\
h^{r+1}\varphi_r^1 &\textrm{for}\quad x \in (a_{i+1},b],
\end{cases}
\end{equation*}
so that the statement holds true.  Let us assume
that~(\ref{statement on integral of phi}) holds for $k-1$.

\noindent We consider three cases. By the definition of $f_i$ we have

\begin{itemize}
\item[(i)] for $x\in[a,a_i)$
\begin{equation*}
\ints_a^x\ints_a^{y_{k-1}}\ldots\ints_a^{y_1}f_i(y_0)\,\dc
y_0\ldots\dc y_{k-2}\,\dc y_{k-1}=0,
\end{equation*}
\item[(ii)] for $x\in[a_i,a_{i+1}]$
\begin{equation*}
\ints_a^x\ints_a^{y_{k-1}}\ldots\ints_a^{y_1}f_i(y_0)\,\dc
y_0\ldots\dc y_{k-2}\,\dc
y_{k-1}=\ints_{a_i}^x\ints_{a_i}^{y_{k-1}}\ldots\ints_{a_i}^{y_1}\varphi_r([a_i,a_{i+1}],y_0)\,\dc
y_0 \ldots \dc y_{k-2}\, \dc y_{k-1}.
\end{equation*}
\item[(iii)] Let  $x\in(a_{i+1},b]$. We write
\begin{multline*}
\ints_a^x\ints_a^{y_{k-1}}\ldots\ints_a^{y_1}f_i(y_0)\,\dc
y_0\ldots\dc y_{k-2}\,\dc
y_{k-1}\\
=\ints_{a_i}^{a_{i+1}}\ints_{a_i}^{y_{k-1}}\ldots\ints_{a_i}^{y_1}\varphi_r([a_i,a_{i+1}],y_0)\,\dc
y_0\ldots \dc y_{k-2}\,
\dc y_{k-1} \\
{\qquad+\ints_{a_{i+1}}^x
\sum\limits_{m=1}^{k-1}\frac{h^{r+m}}{(k-1-m)!}\varphi_r^m (y_{k-1}-
a_{i+1})^{k-1-m}\, \dc y_{k-1}}.
\end{multline*}
The last term in the equation above follows from the inductive
assumption. Integrating the last term and using Fact~\ref{fact
B-spline 1} with $m=k$ to the first term, we get that
\begin{multline*}
\ints_a^x\ints_a^{y_{k-1}}\ldots\ints_a^{y_1}f_i(y)\,\dc y\ldots\dc
y_{k-2}\,\dc
y_{k-1}=h^{r+k}\varphi_r^{k}+\sum\limits_{m=1}^{k-1}\frac{h^{r+m}}{(k-m)!}\varphi_r^m
(x- a_{i+1})^{k-m}.
\end{multline*}
\end{itemize}
This ends the inductive proof of~(\ref{statement on integral of
phi}). To prove  Lemma \ref{lemat function f_i}, it is sufficient to
take  $x=b$ in~(\ref{statement on integral of phi}).
\begin{flushright}
{\large{\qed}}%{\large $\blacksquare $ }
\end{flushright}
\end{pf}

\noindent We are ready to prove Theorem \ref{theorem lower bounds}.

\noindent \begin{pf*}{\textbf{Proof of Theorem \protect\ref{theorem
lower bounds}}} We first  prove a  lower bound on the
$\varepsilon$-complexity in the  quantum setting. Consider the
subclass $\mathcal{G}^r_1$ of $\mathcal{G}^r$. Let $\phi$ be any
quantum algorithm solving problem~(\ref{general problem}) such that
$e^{\rm quant}(\phi,\mathcal{G}^r)\leq \varepsilon$ (this yields
that $e^{\rm quant}(\phi,\mathcal{G}^r_1)\leq \varepsilon$). We
shall prove that ${\rm cost}^\quant(\phi)=\Omega\left(
\left(1/\varepsilon\right)^{1/(r+1)}\right)$.

Note that for any function $g\in\mathcal{G}^r_1$ the
problem~(\ref{general problem}) reduces to the computation of the
\mbox{$k$-fold}
 integral
\begin{equation}
u(x)=\sum_{j=0}^{k-1}\frac{u_a^{j}}{j!}(x-a)^j+\int\limits_{a}^{x}\int\limits_{a}^{t_{k-1}}\ldots
\int\limits_{a}^{t_{1}}g\left( t_0\right) \dc t_0\,\dc t_{1}\ldots
\dc t_{k-1} ,\qquad x\in [a,b].
 \label{problem integral}
\end{equation}
We prove  the lower bound by showing that the solution of a finite
number of problems~(\ref{problem integral}) with suitable right-hand
side functions leads to the solution of the mean value problem.

Let $n\in\mathbb{N}$. We divide the interval $[a,b]$ into $2kn$ parts,
with  points $a_i=a+i\bar{h}$, where $i=0,1,\ldots,2kn$ and
$\bar{h}:=(b-a)/(2kn)$. Let $f_i$ be the functions from~(\ref{function
f_i}) with $n:=2kn$ and $h:=\bar{h}$.

Let $\lambda_i$ be arbitrary numbers with $|\lambda_i|\leqslant 1$ for
$i=0,\ldots,kn-1$, and let $\mathcal{X}$ be an injective mapping
$\mathcal{X}:\{0,1,\ldots, kn-1\}\rightarrow \{0,1,\ldots,2kn-1\}$ .
Then obviously $\sum\limits_{i=0}^{kn-1}\lambda_if_{\mathcal{X}(i)}\in
\mathcal{G}^r_1$.

Consider $k$  such mappings $\mathcal{X}_j$  and numbers $c_j$ to be
specified later on, $j=0,1,\ldots,k-1$. From Lemma \ref{lemat function
f_i}, we have that
\begin{eqnarray*}
S&=&\sum_{j=0}^{k-1}c_j\ints_a^b\ints_a^{y_{k-1}}\ldots\ints_a^{y_1}\sum_{i=0}^{kn-1}\lambda_if_{\mathcal{X}_j(i)}(y_0)
\dc y_0\, \dc y_1\ldots \dc y_{k-1}\\
&=&\sum_{j=0}^{k-1}c_j\sum_{i=0}^{kn-1}\lambda_i\bar{h}^{r+k}\sum_{m=1}^k\frac{1}{(k-m)!}\varphi_r^m
(2kn-\mathcal{X}_j(i)-1)^{k-m}.
\end{eqnarray*}
After changing the summation order, we get that
\begin{eqnarray*}
S&=&\bar{h}^{r+1}\sum_{i=0}^{kn-1}\lambda_i\sum_{m=1}^{k}(b-a)^{k-m}\bar{h}^{m-1}\frac{1}{(k-m)!}\varphi_r^m
\sum_{j=0}^{k-1}c_j\left(1-\frac{\mathcal{X}_j(i)+1}{2kn}\right)^{k-m}.
\end{eqnarray*}
We split the sum indexed by $m$ into two parts. We have
\begin{multline*}
S=\bar{h}^{r+1}\sum_{i=0}^{kn-1}\lambda_i(b-a)^{k-1}\frac{1}{(k-1)!}\varphi_r^1
\sum_{j=0}^{k-1}c_j\left(1-\frac{\mathcal{X}_j(i)+1}{2kn}\right)^{k-1}{}\\
{}+{}\bar{h}^{r+1}\sum_{i=0}^{kn-1}\lambda_i\sum_{m=2}^k(b-a)^{k-m}\bar{h}^{m-1}\frac{1}{(k-m)!}\varphi_r^m
 \sum_{j=0}^{k-1}c_j\left(1-\frac{\mathcal{X}_j(i)+1}{2kn}\right)^{k-m}.
\end{multline*}

\noindent We now show that there exist  mappings $\mathcal{X}_j$ and
numbers $c_j$  such   that for any $i$
\begin{equation}
 \sum_{j=0}^{k-1}
c_j\left(1-\frac{\mathcal{X}_j(i)+1}{2kn}\right)^{k-1}=1,
\label{sum =1}
\end{equation}
 and for $m=2,\ldots,k$,
\begin{equation}
 \sum_{j=0}^{k-1}
c_j\left(1-\frac{\mathcal{X}_j(i)+1}{2kn}\right)^{k-m}=0.
\label{sums = 0 for m=2,...,k}
\end{equation}
We now define  the functions $\mathcal{X}_j$ for  $j=0,\ldots,k-1$.
Let
\begin{equation}
\mathcal{X}_j(i)=n(k-j)+i-1,\qquad i=0,\ldots,kn-1.
\end{equation}
Then obviously $\mathcal{X}_j(i)\in\{0,\ldots,2kn-1\}$. For such
$\mathcal{X}_j$ and $m=1,\ldots,k$, we get that
\begin{align*}
\sum_{j=0}^{k-1}
c_j\left(1-\frac{\mathcal{X}_j(i)+1}{2kn}\right)^{k-m} &=
\sum_{j=0}^{k-1} c_j\left(1-\frac{n(k-j)+i}{2kn}\right)^{k-m}\\
&=\sum_{j=0}^{k-1}
c_j\left(\frac{1}{2}+\frac{j}{2k}-\frac{i}{2kn}\right)^{k-m}.
\end{align*}
\noindent Let now
\begin{equation}
\displaystyle w(x)=\sum\limits_{j=0}^{k-1}
c_j\left(\frac{1}{2}+\frac{j}{2k}-x\right)^{k-1} \label{w(x)}
\end{equation}
be a polynomial of $x$, $x\in \mathbb{R}$. We select $c_j$,
$j=0,\ldots, k-1$, to be numbers such that
\begin{equation}
w(x)\equiv 1. \label{w(x) =1}
\end{equation}
The existence of such numbers follows from the fact that (\ref{w(x)
=1}) defines a system of linear equations with unknown variables $c_j$
and with a transpose Vandermonde matrix. The solution (numbers $c_j$)
of this equation is independent of $n$. From~(\ref{w(x) =1}) we have
that
\begin{equation}
w^{(m)}(x)\equiv 0,
\end{equation}
for $m=1,\ldots,k-1$. Hence
\begin{equation}
\sum\limits_{j=0}^{k-1}
c_j\left(\frac{1}{2}+\frac{j}{2k}-x\right)^{k-m}\equiv 0
\label{w_m(x)}
\end{equation}
for $m=2,\ldots,k$. Taking $x={i}/{(2kn)}$ in (\ref{w(x) =1})
and~(\ref{w_m(x)}) we have the desired property
\begin{equation}
\sum_{j=0}^{k-1}
c_j\left(\frac{1}{2}+\frac{j}{2k}-\frac{i}{2kn}\right)^{k-m}=\begin{cases}
1 & \textrm{for} \quad m=1\\
0 & \textrm{for} \quad m=2,\ldots,k-1\\
\end{cases},
\end{equation}
for all $i=0,1,\ldots,kn-1$. Consequently,~(\ref{sum =1})
and~(\ref{sums = 0 for m=2,...,k}) hold true, and
\begin{multline}
S=\sum_{j=0}^{k-1}c_j\ints_a^b\ints_a^{y_{k-1}}\ldots\ints_a^{y_1}\sum_{i=0}^{kn-1}\lambda_if_{\mathcal{X}_j(i)}(y_0)
\dc y_0\, \dc y_1\ldots \dc
y_{k-1}\\={}\bar{h}^{r+1}(b-a)^{k-1}\frac{1}{(k-1)!}\varphi_r^1
\sum_{i=0}^{kn-1}\lambda_i
\\={}
(kn)^{-r}(b-a)^{k+r}2^{-r-1}\frac{1}{(k-1)!}\varphi_r^1\left(\frac{1}{kn}\sum_{i=0}^{kn-1}\lambda_i\right).
\label{eq: lambda + integrals}
\end{multline}
Denoting
\begin{equation}
I_j=\ints_a^b\ints_a^{y_{k-1}}\ldots\ints_a^{y_1}\sum_{i=0}^{kn-1}\lambda_if_{\mathcal{X}_j(i)}(y_0)
\dc y_0\ldots \dc y_{k-1},
 \label{I_j}
\end{equation}
we have from~(\ref{eq: lambda + integrals}) that
\begin{equation}
\frac{1}{kn}\sum_{i=0}^{kn-1}\lambda_i=(kn)^{r}(b-a)^{-(k+r)}2^{r+1}(k-1)!\frac{1}{\varphi_r^1}\sum_{j=0}^{k-1}c_jI_j.
\label{median of lambda_i}
\end{equation}

For each $j$, we now use the algorithm $\phi$ to compute an
approximation $A_j$ to $I_j$ with  error at most $\varepsilon$ and
probability at least $3/4$. This is done  with cost equal to ${\rm
cost}^\quant(\phi)$. The algorithm $\phi$ approximates the solution
$u(x)$ of problem~(\ref{general problem}), in particular it gives us
an approximation of $u(b)$. Sience the function $u(x)$
satisfies~(\ref{problem integral}) the algorithm $\phi$ can be used to
approximate the $k$-fold integral. Repeating the algorithm
$\Theta(\log k)$ times and computing the median, we improve the
probability of success to be at least $(3/4)^{1/k}$. This is achieved
with  cost equal to $\Theta\left(\log k \cdot {\rm cost
}^\quant(\phi)\right)$.

\noindent Denote $C=(b-a)^{-(k+r)}2^{r+1}(k-1)!/{\varphi_r^1}$. It
follows from~(\ref{median of lambda_i}) that the mean of
$\lambda_i$'s, $1/(kn)\sum\limits_{i=0}^{kn-1}\lambda_i$, is
approximated by $(kn)^rC\sum\limits_{j=0}^{k-1}c_jA_j$ with error
\begin{equation}
\left|
\frac{1}{kn}\sum_{i=0}^{kn-1}\lambda_i-(kn)^rC\sum_{j=0}^{k-1}c_jA_j\right|\leq
\varepsilon_1, \label{mean_value problem}
\end{equation}
 and
probability at least $3/4$, where
$\varepsilon_1:=(kn)^rC\sum\limits_{j=0}^{k-1}|c_j|\varepsilon$. The
cost of doing this  is
 $O\left(k\log k \cdot{\rm cost }^\quant(\phi) \right)$ . We now use lower
complexity bound of Nayak and Wu for computing the mean of $kn$
numbers, see~\cite{Nayak&Wu99}. They showed that the cost of any
algorithm for computing  the mean must be $\Omega
(\min\{kn,1/\varepsilon_1\})$ quantum queries for
$\lambda_0,\ldots,\lambda_{kn-1}$. Taking
$n=\Theta\left((1/\varepsilon)^{1/(r+1)}\right)$ we have  that
$\Omega(\min\{kn,1/\varepsilon_1\})=\Omega\left((1/\varepsilon)^{1/(r+1)}\right)$,
which gives us the desired lower bound. (Note that a query for
$\lambda_0,\ldots,\lambda_{kn-1}$ is also a query for $g$ of the form
$g(x)=\sum\limits_{i=0}^{kn-1}\lambda_if_{\mathcal{X}(i)}(x)$).

In the randomized setting we proceed in a similar way. The difference
appears only in the part connected with  computing the mean value by
randomized algorithm. Algorithm $\phi$ gives us an approximation $A_j$
of each integral $I_j$, for $j=0,\ldots,k-1$, such that
\begin{equation}
\bigg(\textbf{E}\left|A_j-I_j\right|^2\bigg)^{1/2}\leq \varepsilon.
\end{equation}
The error of computing the mean value of $\lambda$'s by using $A_j$ is
\begin{equation}
\left(\textbf{E}\left|
\frac{1}{kn}\sum_{i=0}^{kn-1}\lambda_i-(kn)^rC\sum_{j=0}^{k-1}c_jA_j\right|^2\right)^{1/2}
=\left(\textbf{E}\left|
(kn)^rC\sum_{j=0}^{k-1}c_j\left(A_j-I_j\right)\right|^2\right)^{1/2}.
\end{equation}
Using the properties of the expectation and since
\begin{equation}
\left| \sum_{j=0}^{k-1}c_j\left(A_j-I_j\right)\right|^2\leq\frac{1}{2}
\sum_{i,j=0}^{k-1}|c_ic_j|\bigg(\left|A_i-I_i\right|^2+\left|A_j-I_j\right|^2\bigg),
\end{equation}
 we get
\begin{multline}
\left(\textbf{E}\left(
\frac{1}{kn}\sum_{i=0}^{kn-1}\lambda_i-(kn)^rC\sum_{j=0}^{k-1}c_jA_j\right)^2\right)^{1/2}\\
\leq (kn)^rC \left(\frac{1}{2}
\sum_{i,j=0}^{k-1}\left|c_ic_j\right|\bigg(\textbf{E}\left|A_i-I_i\right|^2+
\textbf{E}\left|A_j-I_j\right|^2\bigg)\right)^{1/2}
 \leq \varepsilon_1,
 \label{randomzied case}
\end{multline}
where
$\displaystyle\varepsilon_1=(kn)^rC\sum_{j=0}^{k-1}|c_j|\varepsilon$.
It was shown in~\cite{Mathe} that any approximation
satisfying~(\ref{randomzied case}) must be based on $\Omega
(\min\{kn,(1/\varepsilon_1)^2\})$ subroutine calls. Hence, to prove
desired bound on the \mbox{$\varepsilon$-complexity} in the randomized
setting it suffices to take
$n=\Theta\left((1/\varepsilon)^{1/(r+1/2)}\right)$. This completes the
proof of Theorem~\ref{theorem lower bounds}.
\end{pf*}
\nopagebreak\begin{flushright}
{\large{\qed}}%{\large $\blacksquare $ }
\end{flushright}%

\noindent From bounds~(\ref{Kacwicz rand}) and~(\ref{Kacewicz quant})
and Theorem~\ref{theorem lower bounds}, we get the following
summarizing corollary.
\begin{col}
The complexity of initial-value problem~(\ref{general problem}) with
an arbitrary $k\geq 1$ satisfies:
\begin{itemize}
\item in the randomized setting
\begin{equation}
\comp^{\rand}(\mathcal{G}^{r},\varepsilon)=O
\left(\left(\frac{1}{\varepsilon}\right)^{1/(r+1/2-\gamma)}\right)
\end{equation}
and
\begin{equation}
\comp^{\rand}(\mathcal{G}^{r},\varepsilon)=\Omega
\left(\left(\frac{1}{\varepsilon}\right)^{1/(r+1/2)}\right),
\end{equation}

\item in the quantum setting

\begin{equation}
\comp^{\quant}(\mathcal{G}^{r},\varepsilon)=O\left(
\left(\frac{1}{\varepsilon}\right)^{1/(r+1-\gamma)}\right)
\end{equation}
and \begin{equation}
\comp^{\quant}(\mathcal{G}^{r},\varepsilon)=\Omega\left(
\left(\frac{1}{\varepsilon}\right)^{1/(r+1)}\right),
\end{equation}
\end{itemize}
where the positive constant $\gamma$ is arbitrary small, and the
constants in the big-O notation depend on $\gamma$.
\end{col}
It follows from the Corollary that the algorithms
from~\cite{Kacewicz05} remain optimal in a subclass of systems arising
from  scalar equations of order $k$, for an arbitrary  $k\geq 1$.

\section{Final remarks}
We showed in this paper   some limitations of randomized and quantum
computing. The intuition that the order $k$ of a differential equation
may help in randomized and quantum setting (which is the case in the
deterministic setting with linear information) turned out not to be
true. Randomized computation gives us a speed-up only by~$1/2$, and
quantum computation  by~$1$ independently of~$k$, which is the same as
in case of the  first order equations. Thus, to optimally solve the
special initial value problem~(\ref{general problem}), we can
transform it to system~(\ref{system of ODEs}) and apply algorithms for
a general system of order~$1$ presented in~\cite{Kacewicz05}.

\vspace{1cm} { \Large \noindent \textbf{Acknowledgement}\linebreak} I
thank  Boles\l aw Kacewicz  for his valuable comments and suggestions.

\end{document}